\documentclass[aps,prm,reprint,amsmath,amssymb,floatfix,superscriptaddress]{revtex4-2}

\makeatletter
\let\frontmatter@footnote@produce\frontmatter@footnote@produce@endnote
\makeatother



\usepackage{graphicx}
\usepackage{physics}
\usepackage{siunitx}
\usepackage{gensymb}
\usepackage{mathtools}
\usepackage{bm}
\usepackage{braket}
\usepackage{amsmath,amsthm,amsfonts}
\usepackage{multirow}
\usepackage{array}
\usepackage{soul}
\usepackage{float}
\usepackage{titlesec}
\usepackage{afterpage}
\usepackage[colorlinks,allcolors=blue,urlcolor=blue]{hyperref}
\usepackage[capitalise]{cleveref}
\usepackage{natbib}

\expandafter\def\expandafter\normalsize\expandafter{%
    \normalsize%
    \setlength\abovedisplayskip{3pt}%
    \setlength\belowdisplayskip{5pt}%
    \setlength\abovedisplayshortskip{-8pt}%
    \setlength\belowdisplayshortskip{2pt}%
}

\begin{document}

\title{Strain-Tunable Spin Relaxation in Germanium from First Principles}

\author{Lauren A. Tan}
\affiliation{Department of Applied Physics and Materials Science, and Department of Physics, \protect\\ California Institute of Technology, Pasadena, California 91125}
\author{Shaelyn Iyer}
\affiliation{Department of Applied Physics and Materials Science, and Department of Physics, \protect\\ California Institute of Technology, Pasadena, California 91125}
\author{Ivan Maliyov}
\affiliation{Department of Applied Physics and Materials Science, and Department of Physics, \protect\\ California Institute of Technology, Pasadena, California 91125}
\author{Jinsoo Park}
\affiliation{Department of Applied Physics and Materials Science, and Department of Physics, \protect\\ California Institute of Technology, Pasadena, California 91125}
\affiliation{Department of Physics, Pohang University of Science and Technology, Pohang 37673, Korea}
\author{Marco Bernardi}
\email{bmarco@caltech.edu}
\affiliation{Department of Applied Physics and Materials Science, and Department of Physics, \protect\\ California Institute of Technology, Pasadena, California 91125}

\begin{abstract} 
Germanium is a leading platform for semiconductor spin qubits and spintronics. Yet its electron and hole spin dynamics remain understood primarily through phenomenological models. Here, we predict electronic transport and spin relaxation in bulk Ge entirely from first principles, combining hybrid-functional band structures with fully relativistic electron-phonon ($e$-ph) interactions. Without empirical parameters, we predict carrier mobilities, velocity-field curves, and electron and hole spin relaxation times in close agreement with experiments over 100$-$400 K. 
By resolving spin-flip scattering by valley and phonon mode, we identify the microscopic mechanisms governing spin relaxation and show that spin and momentum relaxation, although both mediated by $e$-ph scattering, are controlled by distinct processes. We further show that compressive biaxial strain enhances the hole spin lifetime by up to two orders of magnitude at 5\% strain, through strain-induced valence-band splitting and suppressed spin mixing. This mechanism is directly relevant to Ge-on-Si devices, where strain provides a practical route to engineering long-lived hole spins for quantum technologies.\\
\end{abstract}

\maketitle


\titlespacing\subsection{0pt}{12pt plus 4pt minus 2pt}{0pt plus 2pt minus 2pt}

Long spin lifetimes are essential for spintronics and quantum information, enabling both the transport of spin-encoded signals~\cite{Jansen2012May, Han2014Oct, Zelezny2018Mar, Veldhorst2015Oct, Hou2019Jul, Balasubramanian2009May} and the implementation of quantum logic~\cite{Scappucci2021Oct, Banszerus2022Jun, Mi2018Mar, Petit2020Apr, Bourassa2020Dec, Whiteley2019May, Noiri2022Jan}. The coupling of atomic vibrations (phonons) to electron spins is an intrinsic decoherence mechanism: because phonons are ever-present, spin-phonon interactions ultimately limit spin coherence and lifetimes, even in isotopically pure, ultraclean samples. 
Spin-orbit coupling (SOC) gives rise to two key spin-phonon relaxation processes~\cite{Zutic2004Apr}: the Elliott-Yafet (EY) mechanism, in which spin flips accompany electron-phonon ($e$-ph) or electron-defect scattering~\cite{Elliott1954Oct, Yafet1963Jan}, and the Dyakonov-Perel mechanism, in which spins precess between scattering events~\cite{Dyakonov1971}. In centrosymmetric crystals, the EY mechanism dominates, and $e$-ph scattering sets an intrinsic spin lifetime that constrains device performance.
\\
\indent
Germanium is a particularly promising platform for spin-based technologies, owing to its high carrier mobility, strong SOC, abundance of spin-zero isotopes, and compatibility with silicon  technology~\cite{Lawrie2020Oct, Hamaya2018Aug, Kuhn2010Oct}. Hole spins in Ge are especially attractive because they provide all-electrical control enabled by strong SOC ~\cite{Bulaev2007Feb} and the valence band structure lacks valley degeneracy --- advantages that have enabled recent demonstrations of high-fidelity hole-spin qubits~\cite{Watzinger2018Sep, Scappucci2021Oct, Hendrickx2021Mar, Wang2024Jul, Secchi2025May, Zhou2025Aug}. Advancing Ge and other materials platforms for spin technologies requires a quantitative, microscopic understanding of spin-phonon coupling that goes beyond phenomenological models.
\\
\indent
First-principles calculations of $e$-ph interactions can predict electronic transport across a wide range of materials by combining density functional theory (DFT) and density functional perturbation theory (DFPT) with the Boltzmann transport equation (BTE)~\cite{Sohier2014Sep, Li2015Aug, Bernardi2016Nov, Zhou2016Nov, Ponce2018Mar, Zhou2018Nov, Park2020Sep, Zhou2021Jul}. 
This framework has recently enabled quantitative treatments of EY spin relaxation~\cite{Park2020Jan} and unified descriptions of the EY and Dyakonov-Perel mechanisms~\cite{Xu2020Jun,Park2022Nov,Lunghi2022Aug}.
\mbox{Germanium, however,} remains largely unexplored with these first-principles techniques. Its spin physics has been treated primarily using semi-empirical models and symmetry analysis
~\cite{Li2012Aug, Song2014Oct, Tang2012Jan}, while first-principles calculations have so far been limited to hole mobility~\cite{Yang2024}. A central challenge is the electronic structure of Ge: semilocal DFT spuriously closes the band gap~\cite{Perdew1983Nov, Hybertsen}, and an accurate description of the band valleys, effective masses, and spin texture requires more computationally demanding hybrid functionals~\cite{Heyd2003May, Garza2016Oct} or GW quasiparticle calculations~\cite{Hybertsen}.
\\
\indent
In this Letter, we study electronic transport and spin relaxation in bulk Ge entirely from first principles, using hybrid functionals and fully relativistic pseudopotentials to describe the band structure and SOC. 
We predict electron and hole mobilities, velocity-field curves, and spin relaxation times in close agreement with experiments from 100 to 400~K. 
To identify the microscopic mechanisms governing momentum and spin relaxation, we analyze scattering contributions from different phonon modes and electronic valleys. We find that momentum and spin relaxation are controlled by distinct phonon modes, showing that spin lifetimes cannot be inferred directly from transport relaxation times.
Our analysis further predicts that compressive biaxial strain --- the strain state of Ge grown on Si --- increases the hole spin lifetime by up to two orders of magnitude through valence-band splitting and suppressed spin mixing. Our results establish a predictive first-principles description of charge and spin dynamics in Ge and identify strain as a practical knob for engineering hole spin lifetimes.
%
\\
\indent
\textit{Methods.}---Both phonon-limited charge transport and spin relaxation are governed by the $e$-ph coupling matrix elements. In centrosymmetric materials, \mbox{Bloch states} with band index $n$ and crystal momentum $\bm{k}$ form Kramers-degenerate pairs, which can be resolved into effective spin states $\sigma\!=\Uparrow,\Downarrow$ that diagonalize $\hat{S}_z$~\cite{Park2020Jan, Elliott1954Oct, Yafet1963Jan, Pientka2012Aug}. The corresponding spin-resolved $e$-ph matrix elements are~\cite{Park2020Jan}
\begin{equation}
g_{mn\nu}^{\sigma'\sigma}(\bm{k},\bm{q})= \sqrt{\frac{\hbar}{2\omega_{\nu\bm{q}}}}\braket{{m\bm{k}+\bm{q},\sigma'} | \Delta \hat{V}_{\nu \bm{q}} | {n\bm{k},\sigma}},
\label{eq:gspin}
\end{equation}
which give the scattering amplitude from an initial Bloch state $\ket{n\bm{k},\sigma}$ to a final state $\ket{m\bm{k}+\bm{q},\sigma'}$ through interaction with a phonon of mode $\nu$ and wave vector $\bm{q}$. 
In the presence of SOC, the Kohn-Sham potential perturbation $\Delta\hat{V}_{\nu\bm{q}}$ is a $2\times2$ matrix in spin space~\cite{Bernardi2016Nov, Park2020Jan}. The spin-conserving matrix elements ($\sigma'=\sigma$) primarily govern momentum scattering and charge transport, and thus determine the carrier mobility, whereas the spin-flip elements ($\sigma'\neq\sigma$) give rise to EY spin relaxation.
\\
\indent 
We compute the carrier mobility by solving the linearized electronic BTE with first-principles $e$-ph interactions~\cite{Zhou2021Jul}. The mobility is computed as
\begin{equation}
    \mu_{\alpha\beta}(T)=\frac{2e}{ \mathcal{N}_{\bm{k}}V_{\rm uc}\,n_c}\sum_{n\bm{k}}\textbf{v}^\alpha_{n\bm{k}}\bm{F}^{\beta}_{n\bm{k}}(T)\left( -\frac{\partial f_{n\bm{k}}}{\partial \varepsilon_{n\bm{k}}}\right),
\end{equation}
where $\alpha$ and $\beta$ denote Cartesian directions, $T$ is the temperature, $n_c$ is the carrier concentration, $\mathcal{N}_{\bm{k}}$ is the number of $\bm{k}$-points, $V_{\rm uc}$ is the unit-cell volume, $\varepsilon_{n\bm{k}}$ and $\textbf{v}_{n\bm{k}}$ are the carrier energy and velocity, respectively, and $f_{n\bm{k}}$ is the Fermi-Dirac distribution. 
The term $\bm{F}_{n\bm{k}}(T)$ characterizes the deviation from equilibrium of the electron distribution and is obtained by solving the BTE iteratively~\cite{Zhou2021Jul}.
\\
\indent
To model high-field transport, we propagate the real-time BTE under an applied electric field until the carrier distribution reaches a steady state~\cite{Maliyov2021}. We then compute the steady-state drift velocity along the field direction as
\begin{equation}
     \langle v_\parallel(\mathbf{E})\rangle =\frac{1}{\mathcal{N}_{\bm{k}}V_{\rm uc}\,n_c}\sum_{n\bm{k}}\tilde{f}_{n\bm{k}}(\mathbf{E})\left(\textbf{v}_{n\bm{k}}\cdot\hat{\textbf{E}}\right),
    \label{eq:vfield}
\end{equation}
where $\tilde{f}_{n\bm{k}}(\mathbf{E})$ is the steady-state carrier distribution at electric field $\mathbf{E}$, and $\hat{\mathbf{E}}$ denotes the field direction. To obtain the velocity-field curves, we increase the field incrementally, initializing each calculation with the converged steady-state distribution from the preceding field value.
\\
\indent
The state-resolved spin relaxation times are computed within lowest-order perturbation theory~\cite{Park2020Jan, Yafet1963Jan}:
\begin{equation}
    \begin{split}
        \frac{1}{\tau_{n\bm{k}\sigma}^{\text{s}}}=& \frac{2\pi}{\hbar}\sum_{m\sigma'\nu\bm{q}}|g_{mn\nu}^{\sigma'\sigma}(\bm{k},\bm{q})|^2 \Big(1-\frac{s_{m\bm{k}+\bm{q}\sigma'}}{s_{n\bm{k}\sigma}}\Big)\\
        & \times [(N_{\nu\bm{q}}+1-f_{m\bm{k}+\bm{q}})\delta(\varepsilon_{n\bm{k}}-\varepsilon_{m\bm{k}+\bm{q}}-\hbar\omega_{\nu\bm{q}})\\
        &+(N_{\nu\bm{q}}+f_{m\bm{k}+\bm{q}})\delta(\varepsilon_{n\bm{k}}-\varepsilon_{m\bm{k}+\bm{q}}+\hbar\omega_{\nu\bm{q}})],
    \end{split}
    \label{eq:ts}
\end{equation}
where $s_{n\bm{k}\sigma}=\braket{n\bm{k}\sigma|\hat{S}_z|n\bm{k}\sigma}$ is the spin expectation value and $N_{\nu\bm{q}}$ are phonon occupations. The spin lifetime \mbox{relevant to experiment} is obtained as the spin-weighted thermal average~\cite{Park2020Jan, Yafet1963Jan}
\begin{equation}
    \tau_s(T) = \left(\frac{\sum_{n\bm{k}\sigma}\,(\tau_{n\bm{k}\sigma}^{\text{s}})^{-1}\, s_{n\bm{k}\sigma}^2 \,(\partial f_{n\bm{k}}/\partial \varepsilon_{n\bm{k}})}{\sum_{n\bm{k}\sigma}\, s_{n\bm{k}\sigma}^2 \,(\partial f_{n\bm{k}}/\partial \varepsilon_{n\bm{k}})}\right)^{-1},
    \label{eq:t-avg}
\end{equation}
evaluated here using the tetrahedron method~\cite{Blochl1994Jun}.
\\

\indent
%
\begin{figure}[t]
\centering
\includegraphics[width=0.93\columnwidth]{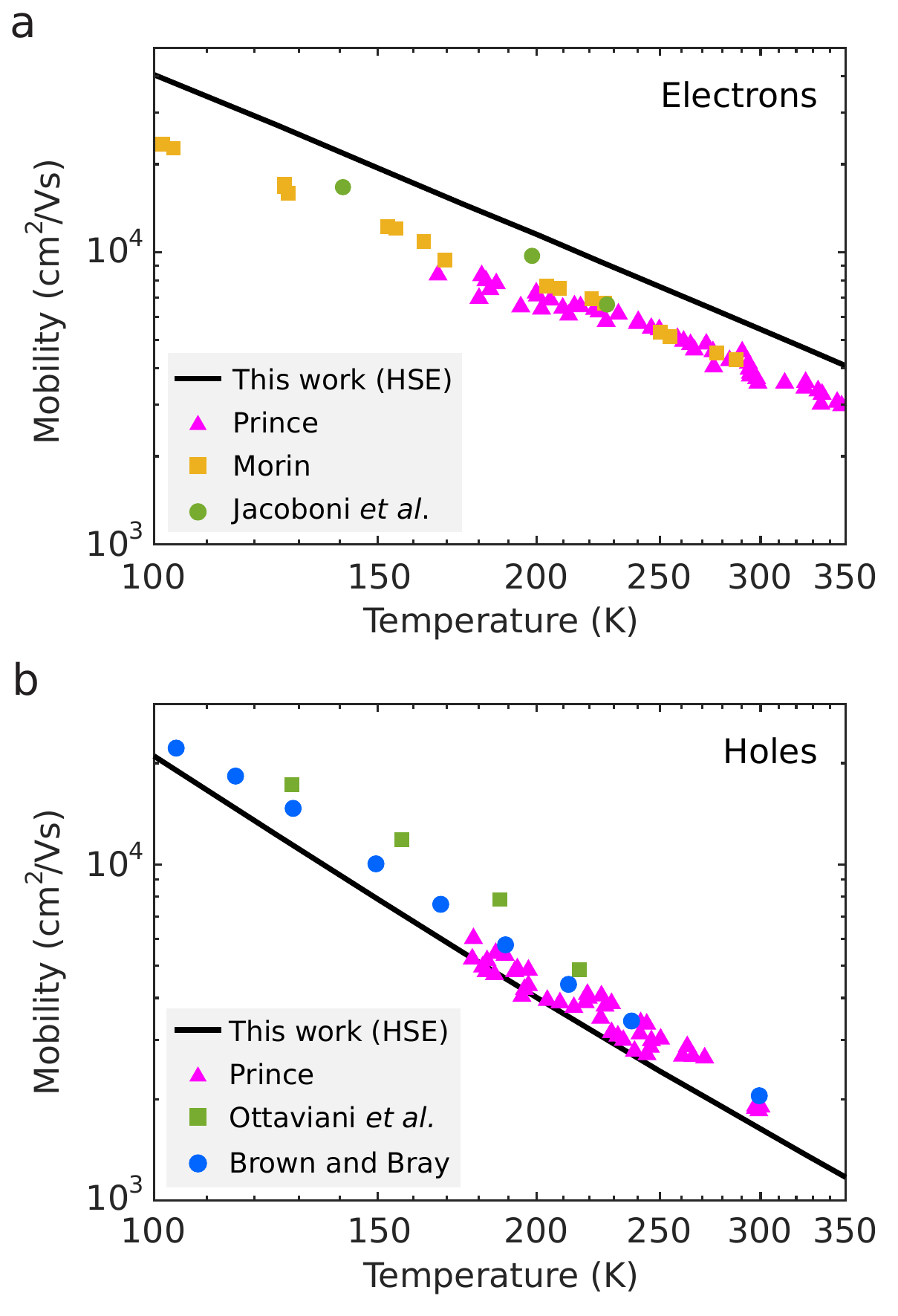}
\caption{Computed carrier mobilities versus temperature for (a) electrons and (b) holes in Ge. Experimental data are from Refs.~\cite{Prince1953Nov,Morin1954Jan,Jacoboni1981Jul} in (a) and Refs.~\cite{Prince1953Nov,Ottaviani1973Jun,Brown1962Sep} in (b).}
\vspace{-5 pt}
\label{fig:fig1mobility}
\end{figure}
\textit{Computational details.}--- We perform DFT calculations using a plane-wave basis in \textsc{Quantum Espresso}~\cite{Giannozzi2009Sep, Giannozzi2017Oct}. The electronic structure is computed using the Heyd-Scuseria-Ernzerhof (HSE06) hybrid exchange-correlation functional~\cite{Heyd2003May}, fully relativistic norm-conserving pseudopotentials~\cite{Theurich2001Jul} from PseudoDojo~\cite{vanSetten2018May}, a converged kinetic energy cutoff of 110~Ry, and a relaxed cubic lattice constant of 5.66~\text{\AA}. These settings accurately reproduce the band gap, effective masses, and valence spin-orbit splitting of Ge~\cite{Cardona1966Feb} (see Supplemental Material (SM)~\cite{supplemental}). 
Phonons and $e$-ph perturbation potentials are computed with DFPT~\cite{Baroni2001Jul} on $16\times16\times16$ $\bm{k}$-point and $8\times8\times8$ and $\bm{q}$-point grids using the local density approximation (LDA), which accurately describes phonon dispersions in Ge~\cite{Baroni2001Jul}. 
We then use \textsc{Perturbo}~\cite{Zhou2021Jul} to compute and interpolate the $e$-ph and spin-flip matrix elements on dense momentum grids, using Wannier functions and spin matrices from \textsc{Wannier90}~\cite{Pizzi2020Jan}, and to compute the transport and spin-relaxation properties. The BTE and spin relaxation calculations use converged $100^3$ $\bm{k}$- and $\bm{q}$-grids.  
For strained Ge, we apply 1--5\% compressive in-plane biaxial strain, mapping the cubic unit cell onto a body-centered tetragonal cell (using a Poisson ratio of $\nu\!=\!0.28$~\cite{Wortman1965Jan}) and relaxing the atomic positions until the forces fall below $10^{-3}$~eV/\text{\AA}.
\\
\indent
\textit{Electronic transport.}---We first validate our band structure and $e$-ph coupling against transport measurements. \Cref{fig:fig1mobility} shows the computed electron and hole mobilities as a function of temperature. 
For electrons, the predicted temperature dependence follows a $T^{-1.8}$ scaling, in close agreement with the measured $T^{-1.7}$ trend over 100--350~K~\cite{Morin1954Jan}. At 300~K, the calculated electron mobility is 5440~cm$^2$V$^{-1}$s$^{-1}$, in good agreement with the experimental value of 3852~cm$^2$V$^{-1}$s$^{-1}$~\cite{Prince1953Nov}. The underlying electronic structure is also accurately captured: our longitudinal conduction band effective mass of $1.65\,m_0$ closely matches the cyclotron-resonance value of $1.64\pm0.03\,m_0$~\cite{Dexter1956Nov}. 
For holes, the calculated mobility agrees with experiment to within $\sim$10\% on average over 100$-$350~K and closely reproduces the measured $T^{-2.2}$ temperature dependence.
At 300~K, the calculated hole mobility is 1626~cm$^2$V$^{-1}$s$^{-1}$, within 12\% of the measured value of 1846~cm$^2$V$^{-1}$s$^{-1}$~\cite{Prince1953Nov}. This difference can be partly attributed to the calculated heavy- and light-hole effective masses of $0.31\,m_0$ and $0.050\,m_0$, respectively, compared with the experimental values of $0.34\,m_0$ and $0.043\,m_0$~\cite{Dexter1956Nov}.
Since our calculations include only $e$-ph scattering, the predicted mobilities provide an upper bound to experiment, as additional scattering from defects and impurities can reduce the mobility, particularly at low temperatures and high doping.
\\
\indent
\begin{figure}
\centering
\includegraphics[width=0.9\columnwidth]{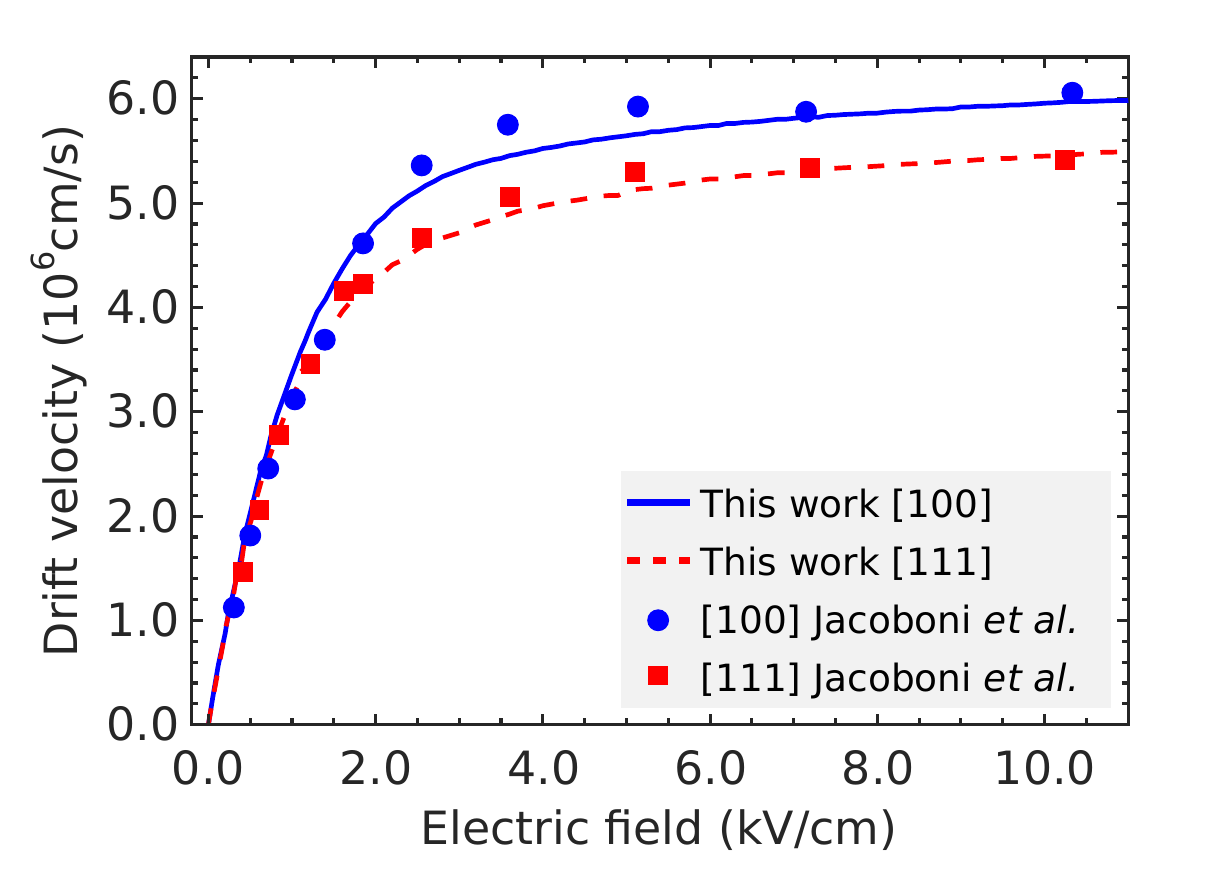}
\caption{Electron velocity-field curves in Ge for electric fields applied along the [100] and [111] directions, computed by solving the real-time BTE and compared with experiment~\cite{Jacoboni1981Jul}.\vspace{-12pt}}
\label{fig:fig2velocity}
\end{figure} 
We further validate our approach by examining high-field transport. \Cref{fig:fig2velocity} shows the electron \mbox{velocity-field} curves for electric fields applied along the [100] and [111] directions. 
The calculated curves closely follow experiment over the full field range up to velocity saturation, with low-field drift velocities within 5\% of the measured values~\cite{Jacoboni1981Jul}. They also reproduce the observed anisotropy, with the [111] velocity saturating below the [100] value.
This behavior originates from the strong anisotropy of the $L$-valley effective masses: the longitudinal mass along [111] is roughly $20$ times larger than the transverse mass, confirming that our band structure accurately captures the valley anisotropy of Ge. 
Together, the mobility and velocity-field results validate our first-principles description of $e$-ph interactions and carrier scattering, providing a robust foundation for the spin-relaxation analysis that follows.
%
\begin{figure}
\centering
\includegraphics[width=0.95\columnwidth]{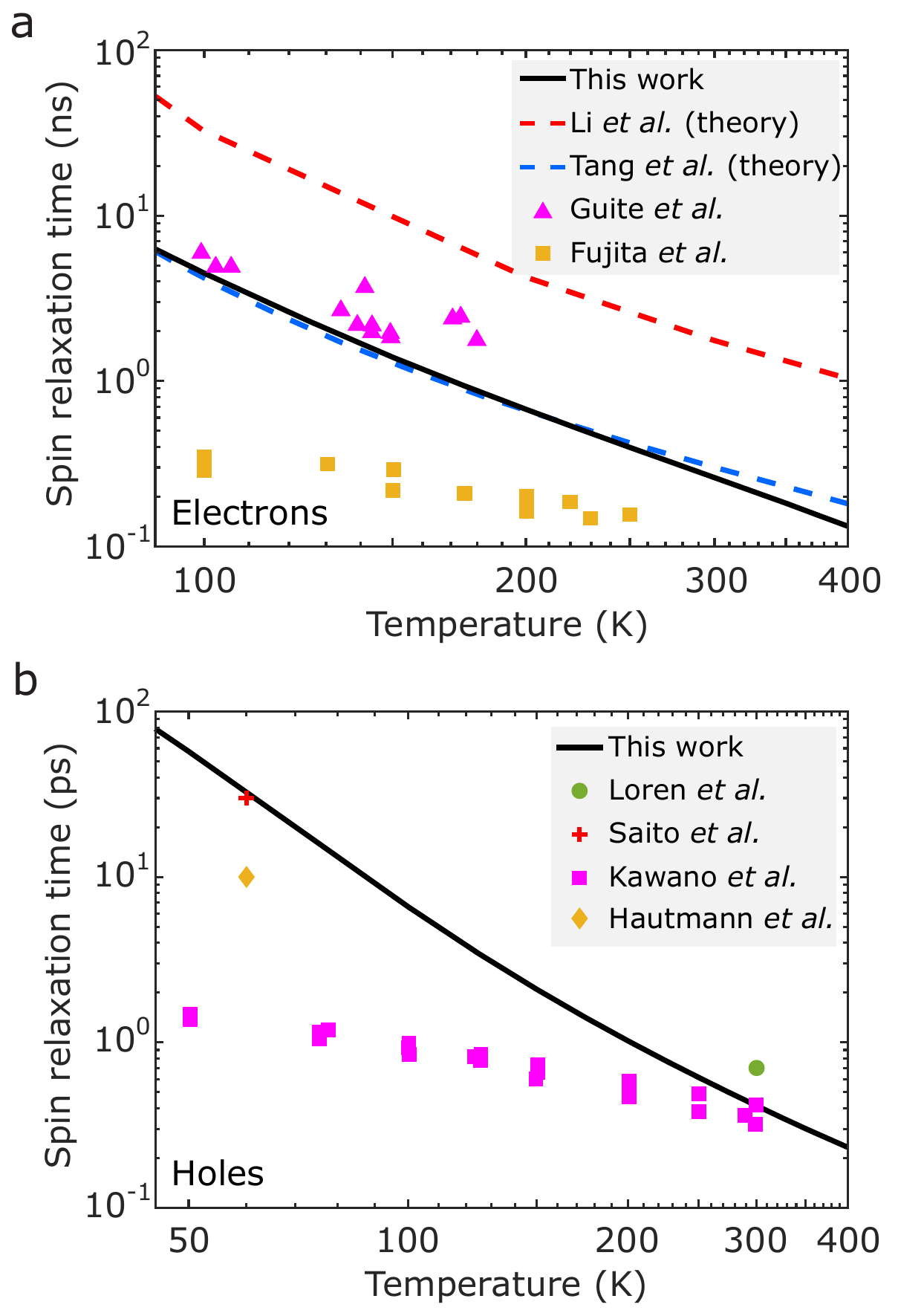}
\caption{Computed spin relaxation times versus temperature for (a) electrons and (b) holes in Ge.  Theory results from Refs.~\cite{Li2012Aug, Tang2012Jan} are shown as dashed lines in (a). Experimental data are from Refs.~\cite{Guite2012Dec,Fujita2017Jul} in (a) and Refs.~\cite{Lange2012,Kawano2017Aug,Loren2011Dec,Saito2011,Hautmann2011} in (b).}\label{fig:figure3srt}
\vspace{-5 pt}
\end{figure}
\\
\indent
\textit{Spin relaxation.}---We now turn to spin relaxation, beginning with electrons in intrinsic Ge. \Cref{fig:figure3srt}(a) shows the computed electron spin relaxation time [Eq.~\eqref{eq:t-avg}] as a function of temperature, together with available experimental data. The calculated spin lifetime is 4.47~ns at 100~K, within 10\% of the 4.88~ns measured in intrinsic Ge by Guite \textit{et al.}~\cite{Guite2012Dec}, and decreases to 0.26~ns at 300~K. 
The agreement with experiments by Guite \textit{et al.}~\cite{Guite2012Dec} is excellent over the entire 100$-$200~K range in their dataset, while the shorter lifetimes reported by Fujita \textit{et al.}~\cite{Fujita2017Jul} are consistent with additional impurity scattering expected at their high doping concentration of $8\times10^{18}$~cm$^{-3}$. 
The temperature dependence exhibits two distinct regimes, scaling as $T^{-3.3}$ below 150~K and $T^{-2.4}$ above, reflecting changes in the phonon modes and intervalley processes that governing spin-flip scattering.
\\
\indent
Previous empirical-model calculations by Li \textit{et al.}~\cite{Li2012Aug} and Tang \textit{et al.}~\cite{Tang2012Jan} reproduce the observed temperature dependence but predict different absolute spin lifetimes because of differences in the empirically fitted coupling parameters. 
In contrast, our first-principles calculations capture both the temperature dependence and the absolute spin lifetimes in close agreement with experiment, without adjustable parameters.
\\
\indent
Strong SOC in the valence bands leads to hole spin relaxation in Ge that is more than two orders of magnitude faster than electron spin relaxation. \Cref{fig:figure3srt}(b) shows the calculated hole spin relaxation times from 50 to 400~K. The temperature dependence exhibits two distinct regimes, scaling as $T^{-2.9}$ below 200~K and $T^{-2.1}$ above. At 300~K, the calculated lifetime is 0.410~ps, in close agreement with the experimental value of 0.316~ps reported by Kawano \textit{et al.}~\cite{Kawano2017Aug}. 
Across different experimental techniques and doping levels, the measured hole spin lifetimes are broadly consistent with our  predictions for intrinsic Ge~\cite{Loren2011Dec,Saito2011,Kawano2017Aug,Hautmann2011}. 
\\
\indent
Optical measurements on bulk Ge~\cite{Loren2011Dec} and spin-injection measurements at a low hole concentration ($n\!=\!3\times10^{17}$~cm$^{-3}$)~\cite{Saito2011} closely agree with our results, whereas the shorter lifetimes observed at higher doping ($n\!=\!1\times10^{18}$~cm$^{-3}$)~\cite{Kawano2017Aug} are consistent with additional impurity scattering. The measured $T_2^*$ values~\cite{Hautmann2011}, which can include additional inhomogeneous dephasing, are also consistent with our predictions. 
Overall, our phonon-limited calculations capture the intrinsic hole spin lifetimes in Ge and provide an upper bound for samples with additional impurity scattering.
\\
\indent
\begin{figure}
\centering
\includegraphics[width=0.95\columnwidth]{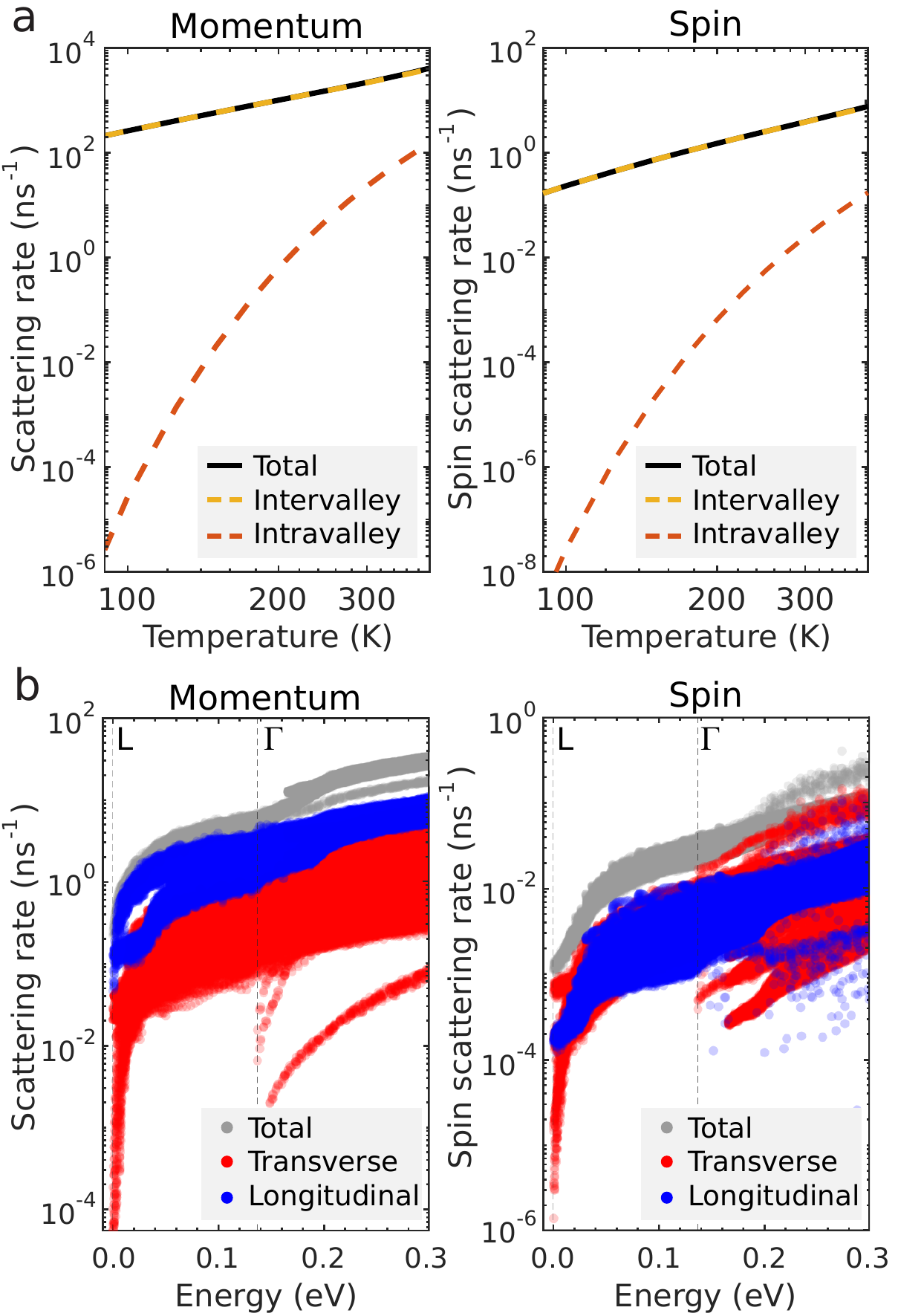}
\caption{(a) Momentum (left) and spin (right) scattering rates for electrons in Ge, defined as the inverse momentum ($\sigma\!=\!\sigma'$) and spin ($\sigma\!\neq\!\sigma'$) relaxation times in Eq.~\eqref{eq:ts}, respectively, resolved into inter- and intravalley contributions as a function of temperature. (b) Same scattering rates at 300~K, resolved into longitudinal and transverse phonon-mode contributions and plotted as a function of energy near the conduction band minimum. The bottoms of the $L$ and $\Gamma$ valleys are indicated by vertical dashed lines, with the $L$-valley minimum taken as the energy zero.}
\label{fig:figure4valley}
\vspace{-5 pt}
\end{figure}
\textit{Spin versus momentum scattering.}---An important question is whether spin and momentum relaxation share the same microscopic mechanisms in Ge, as is often assumed when spin lifetimes are inferred from transport within the EY picture~\cite{Elliott1954Oct}. 
To address this question, we analyze the momentum and spin scattering rates, obtained respectively by setting $\sigma\!=\!\sigma'$ and $\sigma\!\ne\!\sigma'$ in Eq.~\eqref{eq:ts}\footnote{The term in parentheses on the right-hand side of Eq.~\eqref{eq:ts} is omitted for momentum scattering, but is important for spin scattering in the presence of strong spin mixing.}, and decompose them into inter- and intravalley contributions by restricting the sum over $\bm{q}$ to the corresponding regions of the Brillouin zone, as well as into longitudinal and transverse phonon-mode contributions.
This analysis, shown in \cref{fig:figure4valley}, reveals that spin and momentum relaxation exhibit distinct trends and arise from different microscopic scattering processes.
\\
\indent
Intervalley scattering between degenerate $L$ valleys dominates both momentum and spin scattering at all temperatures, while intravalley scattering remains a minor contribution. 
This result is consistent with previous symmetry analyses based on empirical models~\cite{Li2012Aug,Tang2012Jan}, which identified intervalley scattering as the dominant spin-relaxation channel from 100 to 400~K.
However, the contributions of different phonon modes to momentum and spin relaxation differ, as illustrated in Fig.~\ref{fig:figure4valley}(b), which compares the corresponding scattering rates for electrons near the conduction band minimum at 300~K. 
Near the band edge, momentum relaxation is dominated by longitudinal phonons, whereas spin relaxation receives comparable contributions from longitudinal and transverse phonons. This behavior mirrors our previous findings in Si~\cite{Park2020Jan}, where transverse phonons likewise make a substantial contribution to spin relaxation, pointing to a common feature of spin-phonon scattering in group-IV semiconductors.
Because the spin-flip and spin-conserving channels weight phonon modes and coupling strengths differently, spin relaxation times cannot be reliably inferred from momentum relaxation times obtained from transport or optical measurements. These findings underscore the need for quantitative, first-principles methods to predict spin relaxation in materials. 
%
\begin{figure}
\centering
\includegraphics[width=0.85\columnwidth]{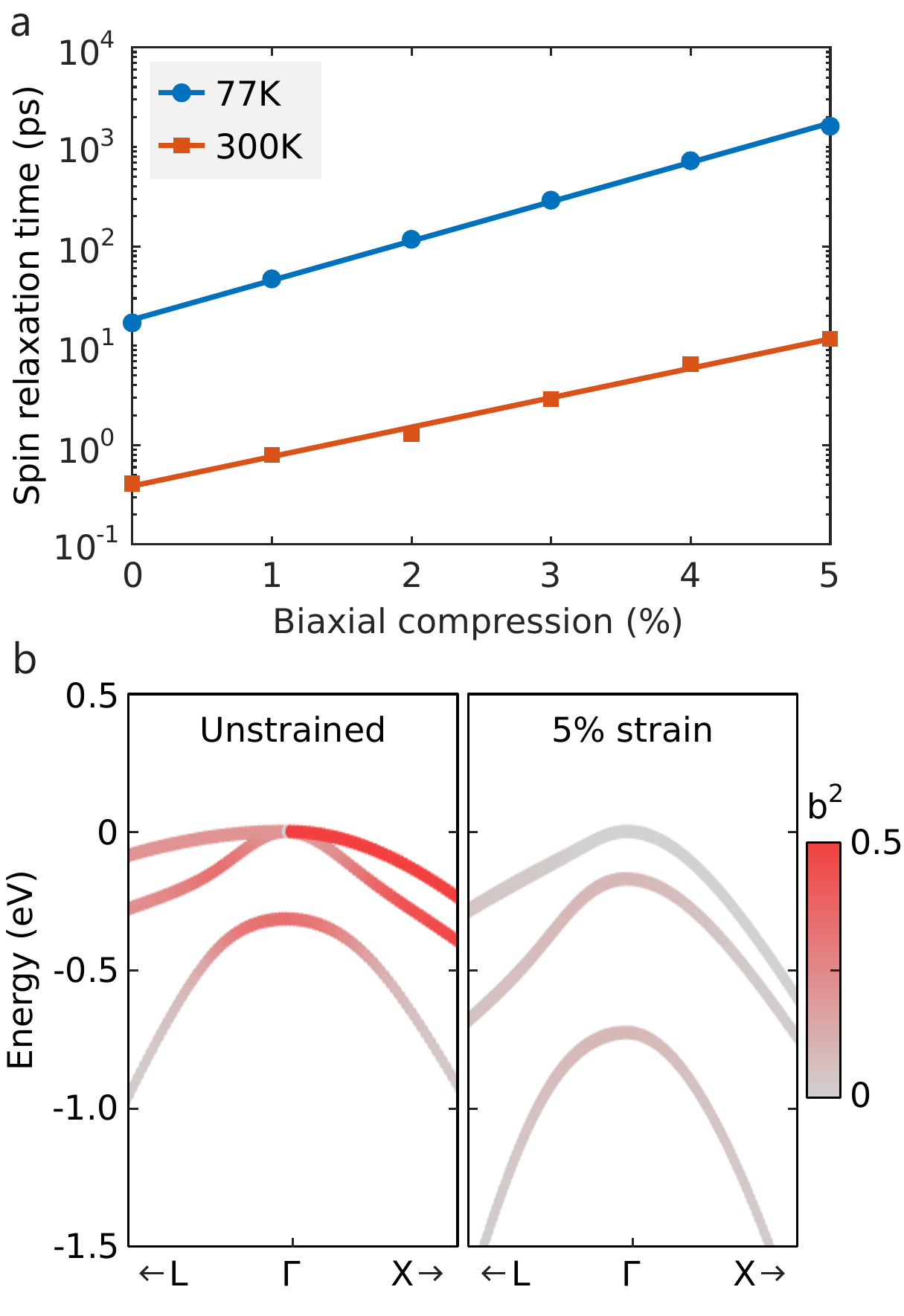}
\caption{(a) Hole spin relaxation time as a function of compressive biaxial strain at 77~K (circles) and 300~K (squares); lines are guides to the eye. (b) Spin-resolved valence bands of unstrained (left) and 5\%-strained (right) Ge, color-coded by the spin-mixing parameter $b^2$, with red indicating strong spin mixing. Energies are referenced to the respective valence band maximum.\vspace{-2pt}}
\label{fig:figure5strain}
\vspace{-5 pt}
\end{figure}
\\
\indent
\textit{Strain effects on hole spin relaxation.}---Strain is a standard knob for band-structure engineering in semiconductors. In Ge, tensile strain drives an indirect-to-direct gap transition and has been studied extensively~\cite{Soref1993Sep, Menendez2004Aug, Chang2007Mar, Mellaerts2021Jul, Geiger2015Jul}. In contrast, the effect of strain on spin relaxation has been examined only using analytical models~\cite{Li2012Aug, Tang2012Jan}, while first-principles studies of strained Ge have focused primarily on its electronic structure~\cite{Tahini2012Apr, Yang2013Oct, Sakata2016Feb, Liu2014Sep}.
\\ 
\indent 
\Cref{fig:figure5strain}(a) shows a central result: under compressive in-plane biaxial strain, the hole spin relaxation time in Ge increases by nearly two orders of magnitude from 0\% to 5\% strain, at both 77~K and 300~K. In contrast, the electron spin relaxation time is essentially unaffected by strain, as we have verified.
The spin-resolved valence bands in \cref{fig:figure5strain}(b) reveal the microscopic origin of this behavior. 
Due to SOC, the Bloch states are mixtures of the two spin states, $|\psi_{n\bm{k}\sigma}\rangle \!=\! a_{n\bm{k}\sigma}\lvert\,\uparrow\,\rangle + b_{n\bm{k}\sigma}\lvert\,\downarrow\,\rangle$, where the spin mixing parameter $b^2_{n\bm{k}\sigma}$ quantifies the deviation from a pure spin state and is related to the spin expectation value by $b^2 \!=\! (1 \!-\! |\langle \sigma_z \rangle|)/2$~\cite{Zutic2004Apr}; therefore, $b^2$ ranges from zero for a pure spin state to 0.5 for maximal spin mixing. 
In unstrained Ge, the heavy- and light-hole bands are degenerate at $\Gamma$ and strongly spin-mixed ($b^2\!\approx\!0.5$), providing efficient spin-flip channels. Compressive strain lifts this degeneracy and pushes the heavy-hole band upward, leaving the states near the valence-band edge energetically isolated and nearly spin-pure. The resulting band separation suppresses interband spin-flip scattering, while the reduced spin mixing weakens the EY matrix elements. 
These two effects act together to produce the large enhancement of the hole spin lifetime. Importantly, this is the strain state of Ge epitaxially grown on Si, where the lattice mismatch imposes approximately 4\% in-plane compression. 
Our results therefore predict substantially longer-lived hole spins in Ge-on-Si heterostructures than in bulk unstrained Ge, with strain providing a practical and directly accessible knob for engineering spin relaxation.
%
\\
\indent
\textit{Conclusion.}---We have developed a parameter-free, first-principles description of electronic transport and spin relaxation in Ge that achieves quantitative agreement with experiments for both electrons and holes. We predict that compressive biaxial strain enhances the hole spin lifetime by up to two orders of magnitude through the combined effects of valence-band splitting and reduced spin mixing. These results establish a microscopic foundation for controlling spin dynamics in Ge through strain engineering.
\\
\indent
Several directions for future work follow naturally. Our present treatment focuses on bulk Ge, where quantum confinement and interface effects are absent. 
Extending these results to nanometer-scale Ge --- particularly ultrathin films and gated quantum dots used for Ge hole-spin qubits~\cite{Terrazos2021Mar, Hendrickx2021Mar} --- requires addressing the Dyakonov-Perel mechanism~\cite{Park2022Nov} and two-phonon spin relaxation processes relevant at low temperatures~\cite{2ph-Si} to bridge bulk predictions with real quantum devices. 
Similarly, extending the calculations to SiGe alloys and Ge/SiGe quantum wells would connect directly to the heterostructures now used in experiments, where alloy disorder and interface scattering add channels beyond the phonon-limited regime studied here. 
Addressing these open questions will advance first-principles spin dynamics toward a predictive framework for quantum device engineering.
\\
\\ 
\noindent
L.T. is supported by the National Science Foundation Graduate Research Fellowship under Grant No. 2139433. J.P. is supported by the Glocal University 30 project (Grant No. 2.0081021.03). This research used resources of the National Energy Research Scientific Computing Center, a DOE Office of Science User Facility supported by the Office of Science of the U.S. Department of Energy under Contract No. DE-AC02-05CH11231 using NERSC award DDR-ERCAP0026831.
\\
\\
\textit{Data availability}---The data that support the findings of this Letter will be made available before publication.
\nocite{Zwerdling1959Apr}
%
%
\\
\bibliography{main}

\end{document}